\documentclass[aps,pra,twocolumn,floatfix,superscriptaddress,showpacs]{revtex4}

\usepackage{graphicx} 

\usepackage{amsmath}

\usepackage{bm}

\usepackage{dcolumn}

\begin{document}

\title{Level-crossing spectroscopy of the 7, 9, and 10D$_{5/2}$ states of $^{133}$Cs 
and validation of relativistic many-body calculations of the polarizabilities
and hyperfine constants}

\author{M.~Auzinsh}

\email{mauzins@latnet.lv}

\affiliation{Department of Physics and Mathematics, University of Latvia,
  Rainis Blvd. 19, Riga LV-1586, Latvia}

\author{K.~Blushs}
\affiliation{Department of Physics and Mathematics, University of Latvia,
  Rainis Blvd. 19, Riga LV-1586, Latvia}

\author{R.~Ferber}
\affiliation{Department of Physics and Mathematics, University of Latvia,
  Rainis Blvd. 19, Riga LV-1586, Latvia}

\author{F.~Gahbauer}
\affiliation{Department of Physics and Mathematics, University of Latvia,
  Rainis Blvd. 19, Riga LV-1586, Latvia}

\author{A.~Jarmola}
\affiliation{Department of Physics and Mathematics, University of Latvia,
  Rainis Blvd. 19, Riga LV-1586, Latvia}

\author{M.~S.~Safronova}

\affiliation {Department of Physics and Astronomy, 223 Sharp Lab, University of Delaware, Newark, Delaware 19716}

\author{ U.~I.~Safronova}

\affiliation{Physics Department, University of Nevada, Reno, Nevada 89557}

\author{M.~Tamanis}
\affiliation{Department of Physics and Mathematics, University of Latvia,
  Rainis Blvd. 19, Riga LV-1586, Latvia}

\pacs{32.10.Dk,32.10.Fn,31.15.Ar,31.25.Jf}

\begin{abstract}
We present an experimental and theoretical investigation of the
polarizabilities and hyperfine constants of 
D$_J$ states in $^{133}$Cs for $J=3/2$ and $J=5/2$.  
New experimental values for the hyperfine constant $A$ are obtained from
level-crossing signals of the (7,9,10)D$_{5/2}$ states of $^{133}$Cs and
precise calculations of the tensor polarizabilities $\alpha_2$.  The
results of relativistic many-body calculations for scalar and tensor 
polarizabilities of the (5-10)D$_{3/2}$ and (5-10)$D_{5/2}$ states are
presented and compared with measured values from the literature.  
Calculated values of the hyperfine constants $A$ for these states are also
presented and checked for consistency with experimental values.

\end{abstract}
\date{\today}

\maketitle

\section{Introduction}

Level-crossing spectroscopy in an electric field has been shown to be a useful
technique to
determine atomic properties.  Already the first experimental
studies of resonant signals at pure electric field crossings of magnetic $\pm
m_F$ components of certain hyperfine (hfs) atomic levels $F$ at nonzero
electric field~\cite{Kha66,Kha68,Sch71} and their further development by
applying two-step laser excitation~\cite{Auz06} demonstrated how this
technique could be used to obtain atomic properties.  The method makes use of
the fact that the electric field values at which magnetic sublevels $m_F$ cross in an
electric field depend on the tensor polarizability $\alpha_2$ and on the hfs constants.
When the electric field is scanned and laser induced fluorescence (LIF) of
definite polarization is observed, these crossings are associated with
resonance behavior in the LIF signals.  
When the separation between crossings is large compared to the widths of the
resonance signals, as in the $n$D$_{3/2}$ states of the $^{133}$Cs  atom
(see Fig.~\ref{lc}a), they lead
to rather well-pronounced resonances in the observed fluorescence.  Moreover, these 
resonances correspond exactly to the level-crossing 
points under appropriate experimental conditions.  Such resonances
were used to measure the tensor polarizabilities $\alpha_2$ in the
$7,9$D$_{3/2}$ states of $^{133}$Cs atoms~\cite{Auz06}, in which the magnetic
dipole coupling
hfs constant $A$ had been previously measured with good precision, and the electric quadrupole hfs
constant $B$ was assumed to be negligibly small~\cite{Ari77}.

Such measurements become more challenging, however, in the case of the
$n$D$_{5/2}$ states of cesium, since there are many closely spaced crossing
points of magnetic $\pm m_F$ components (see Fig.~\ref{lc},b-d).  As a result, the 
level-crossing
signals overlap and no longer contain discernable resonances.  In this case,
reliable values for atomic properties can be extracted only by means of a very
detailed and accurate theoretical description of the observed
electric field dependence  of the signals as a function of atomic properties 
and experimental conditions.  Such theoretical descriptions have been
developed and tested in connection with the $n$D$_{3/2}$ states~\cite{Auz06}.
Nevertheless, the level-crossing technique cannot be used at this time to
improve the knowledge of the tensor polarizabilities $\alpha_2$ of the
$n$D$_{5/2}$ states because the extant measurements of the hfs constant $A$
contain uncertainties on the order of 30\%.  The small hyperfine interaction,
especially for $n>7$, makes them difficult to measure~\cite{Hog75,Sva75}.  

The first value of the hfs constant $A$ of the $n$D$_{5/2}$ states of $^{133}$Cs was obtained with measurements
of the widths of optical double resonance (ODR)
signals in the Paschen-Back region.  The results for the
9D$_{5/2}$ and 10D$_{5/2}$ states were $-0.5(2)$ MHz and $-0.4(2)$ MHz,
respectively~\cite{Sva73}.  These values were improved through level-crossing
spectroscopy in magnetic fields, yielding $-0.40(15)$ and $-0.30(10)$
MHz~\cite{Hog75}.  The 
authors combined these data with previous ODR measurements~\cite{Sva75} and
presented the weighted average as $-0.45(10)$ and $-0.35(10)$ MHz.  They
concluded that the quadrupole interaction can be completely ignored when
fitting the experimental data.  For the 7D$_{5/2}$ state, Bulos \emph{et al.}
estimated the $A$ value to be $-1.7(2)$ MHz from the ODR signal
width~\cite{Bul76}.  The drawback of the ODR experiments on the
7,9,10D$_{5/2}$ signals is that it is necessary to use indirect cascade
transitions to observe the $n$D$_{5/2}$ signals because of the presence of
scattered light at the $n$D$_{5/2} \longrightarrow$ 6P$_{3/2}$ fluorescence 
transition~\cite{Sva75,Bul76}

The tensor polarizabilities
$\alpha_2$ for the $n$D$_{5/2}$ states in cesium under discussion are known
with far greater precision.  For the 10D$_{5/2}$ state, $\alpha_2$ has been
measured by Xia and
coworkers to a very high precision of about 0.3\% at $6815(20)\times 10^3$
a.u.~\cite{Xia97}.  For the 9D$_{5/2}$ state the $\alpha_2$ value is measured
with ca. 5\% accuracy at $2650(140)\times 10^3$ a.u. by means of level crossing
spectroscopy in combined electric and magnetic fields~\cite{Hog75,Fre77}.  For the 7D$_{5/2}$
state, there exists a measured value of $129(4)\times 10^3$ a.u. presented in
Ref.~\cite{Wes87}, which, however, should be verified, because it differs
significantly from the theoretical value of $140 \times 10^3$
a.u.~\cite{Wij94}.  Furthermore, a more recent measurement of $\alpha_2$ for
the 7D$_{3/2}$ state~\cite{Auz06} was closer to the theoretical estimate
of~\cite{Wij94} than the measurement of~\cite{Wes87}.

The situation with the electronic structure calculations is similar to the
experimental situation.  Rather
good accuracy has been achieved for theoretical estimates of the tensor
polarizability $\alpha_2$, as can be seen from the fact that the calculations
of \cite{Wij94} for $^{133}$Cs agree with very accurate experimental data for
the (10-13)D$_{3/2,5/2}$ states~\cite{Xia97}.  Despite this precision for the
polarizability, the estimates of the hfs constants are poor and can hardly be
evaluated reliably, for reasons that will be discussed below.  Therefore,
there is a need for more accurate values for the hfs constants of the 
$n$D$_{5/2}$ states.

In order to determine the hfs constants 
$A$ from our measurements of $m_F$ sublevel crossing signals in the
7,9,10D$_{5/2}$ states of cesium, we used the following approach.  
We fit the measured signals with calculated curves derived from simulations,
which had been developed and tested
in~\cite{Auz06}.  With the tensor
polarizability $\alpha_2$ fixed, these fits yielded the hfs
constant $A$.  To choose the proper value for $\alpha_2$, we performed
 an all-order relativistic many-body calculation.

Section II contains a description of the experiment, followed by a discussion of the
simulations used to describe the measured signals.  The all-order relativistic
many-body calculations that provided the values for the tensor
polarizabilities $\alpha_2$ are described in section III, and the values for $\alpha_2$
obtained from these calculations are compared with earlier experimentally 
measured values.  
In section IV we discuss the
application of these calculations to estimating the hfs constants and compare
them to the results of previous experiments.  In section V we show how to use
our experimental results from section II and the calculated tensor polarizabilities from section III to 
estimate new values for the hfs constant $A$.

\section{Experiment and Description of Signals}
\subsection{Method}
\begin{figure}[tbp]
\centering
\includegraphics[width=0.4\textwidth]{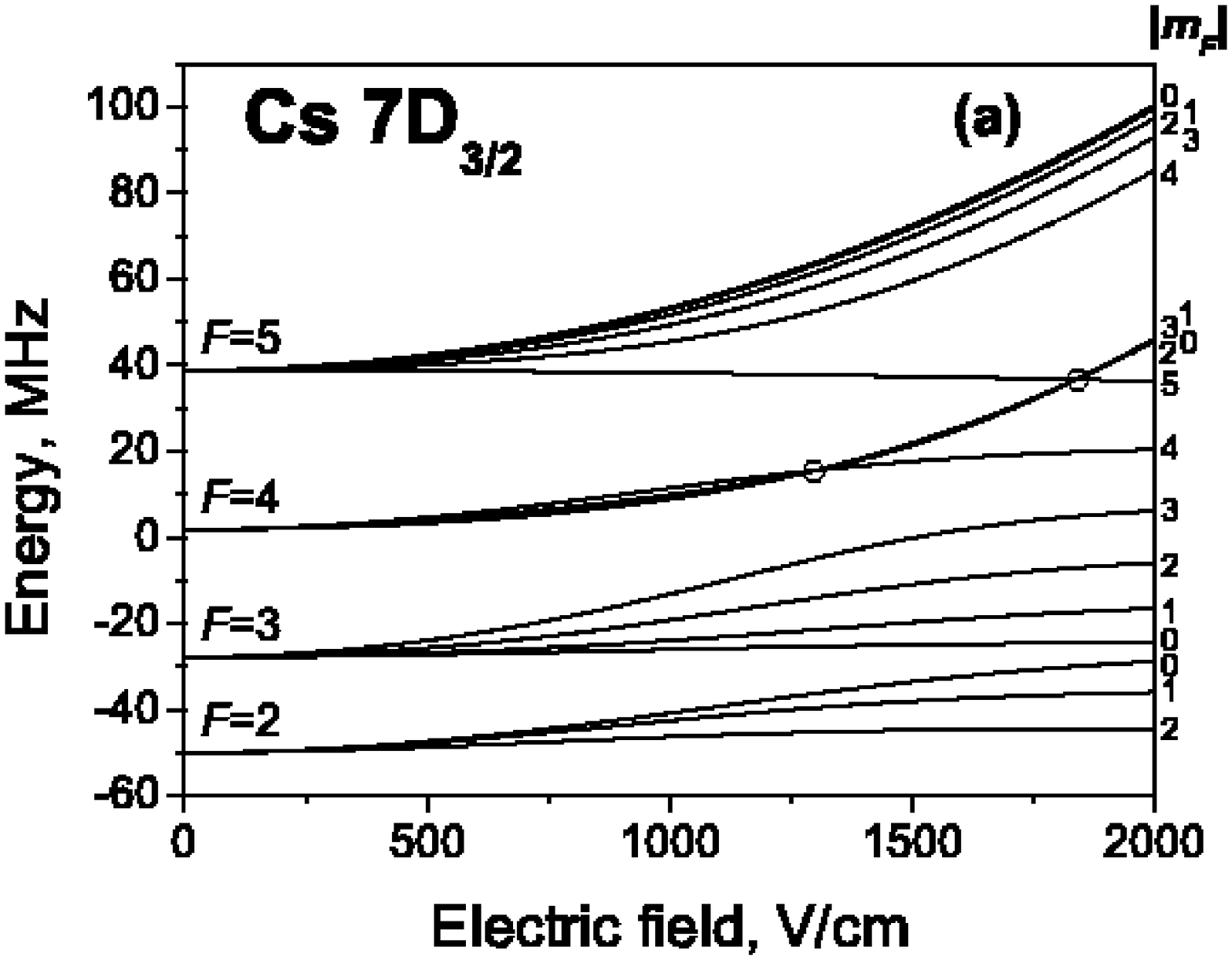}\hfill 
\includegraphics[width=0.4\textwidth]{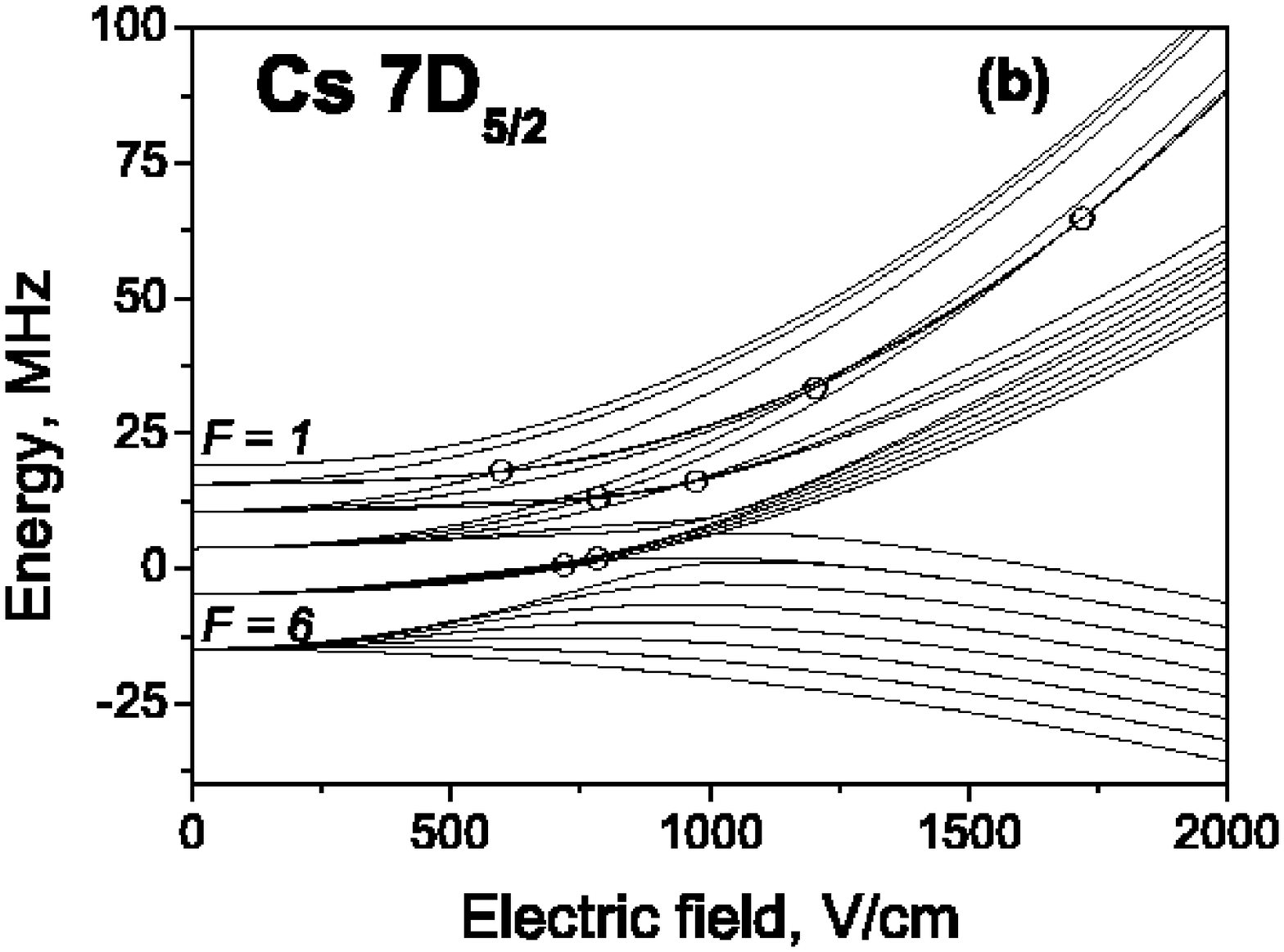}\hfill 
\includegraphics[width=0.4\textwidth]{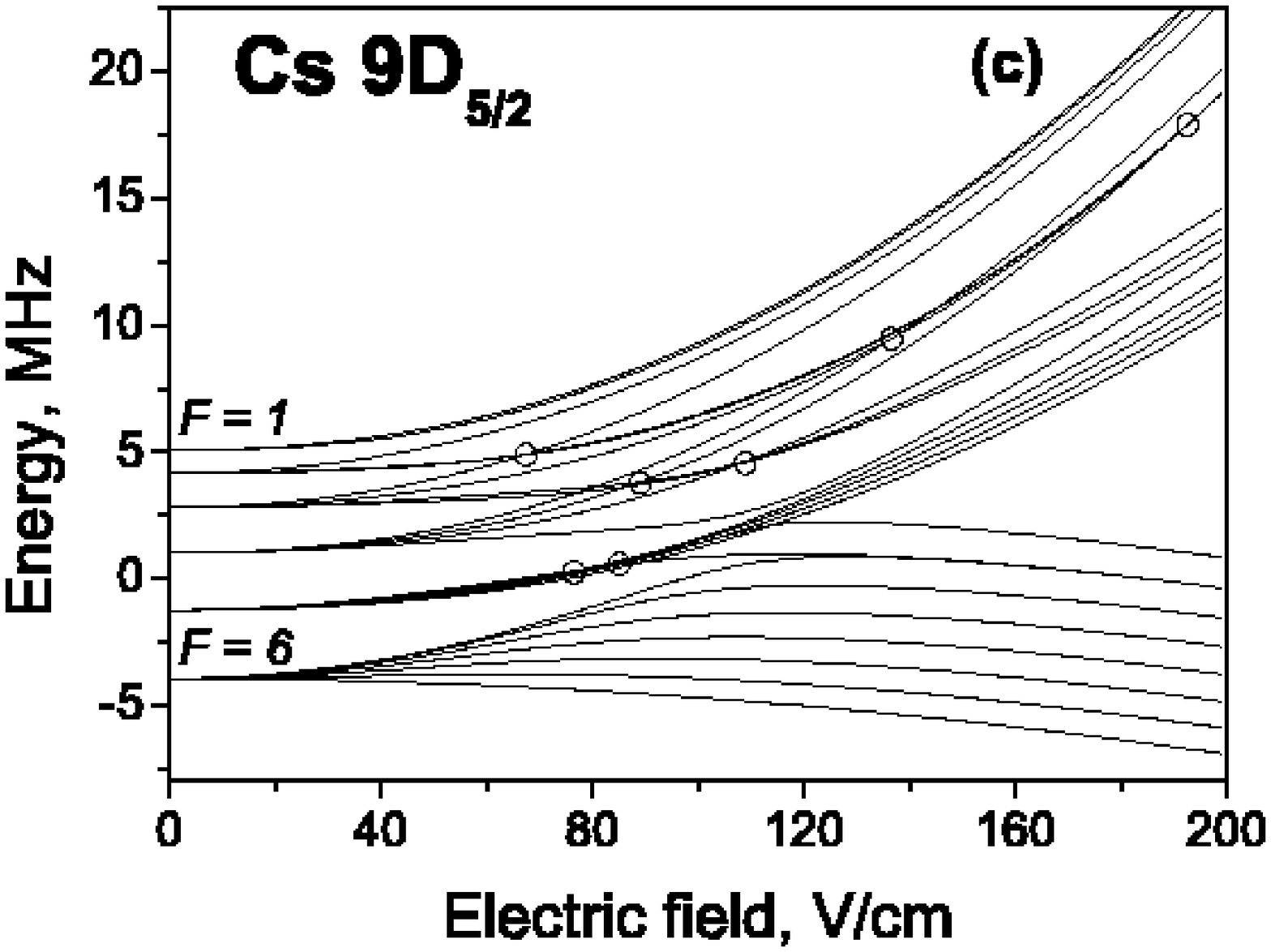}\hfill 
\includegraphics[width=0.4\textwidth]{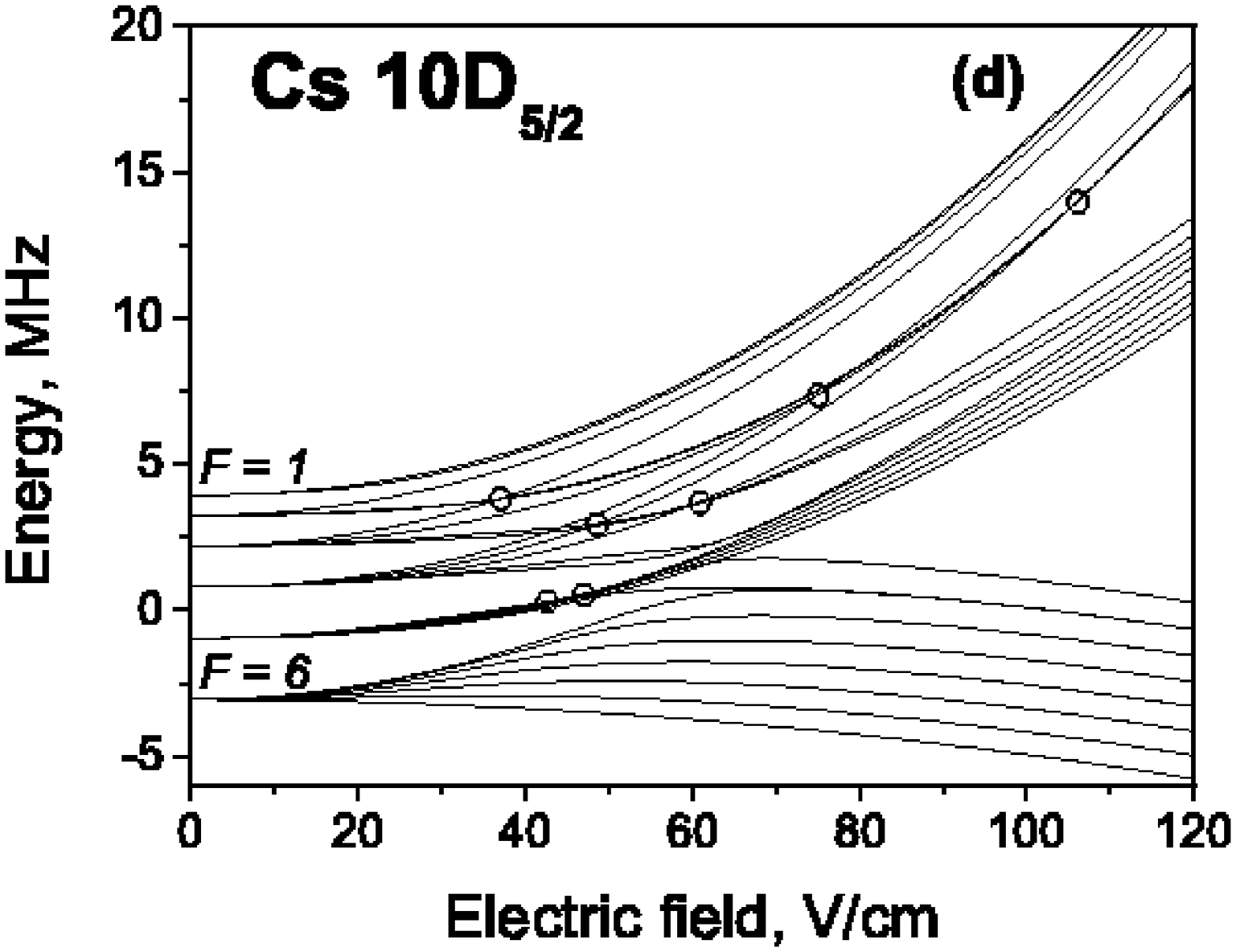}\hfill 
\caption{Hyperfine level-splitting diagram in an external electric field for
  the (a) 7D$_{3/2}$, (b) 7D$_{5/2}$, (c) 9D$_{5/2}$, and (d) 10D$_{5/2}$ states of Cs.  Circled points
  indicate level crossings with $\Delta m_F=\pm 2$.}
\label{lc}
\end{figure}

The premise of level crossing spectroscopy is that the spatial intensity distribution and polarization of
the laser induced fluorescence produced when an atom is excited depends
on the coherences between different magnetic sublevels $m_F$ of hyperfine
levels $F$.  Such coherences are destroyed when the degeneracy between
different sublevels is broken in an electric field.  However, in the case of linear polarization, they can be
restored when sublevels with $\Delta m_F=\pm 2$ cross at certain electric field values.
Figure~\ref{lc} shows the hyperfine level-splitting diagram in an external
electric field for the 7D$_{3/2}$ and 7D$_{5/2}$ states of cesium.  
This
diagram is calculated by diagonalizing the Hamiltonian, which includes the
hyperfine and Stark interactions, in an uncoupled basis~\cite{Aleksandrov93}.

When applying the method of level crossing spectroscopy to the study of the $n$D$_{5/2}$
states of cesium, one encounters two difficulties not present in the case of
the $n$D$_{3/2}$ states.  The first difficulty is that the $n$D$_{5/2}$
hyperfine manifold contains seven level crossings with $\Delta m_F = \pm 2$,
whereas the $n$D$_{3/2}$
manifold contains only two (see Fig.~\ref{lc}).  
The large number of level crossings in the $n$D$_{5/2}$ state wash out
the sharp resonances that could be observed in the $n$D$_{3/2}$ state.  
 
\begin{figure}[tbp]
\includegraphics[width=0.45\textwidth]{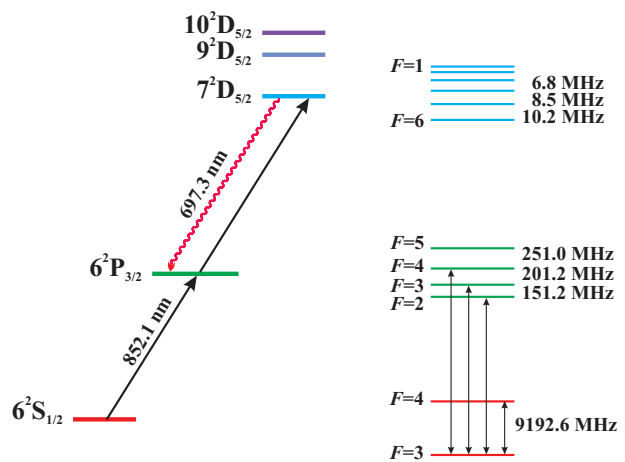} 
\caption{Level excitation scheme.}
\label{excitation}
\end{figure}

The second difficulty is that in the case of the $n$D$_{5/2}$ states,
after the two-step excitation 6S$_{1/2} \longrightarrow 6$P$_{3/2}$
$\longrightarrow n$D$_{5/2}$ (see Fig.~\ref{excitation}), it is necessary 
to observe the fluorescence from the $n$D$_{5/2} \longrightarrow 
6$P$_{3/2}$ transition.  Thus, scattered light 
from the exciting laser constitutes a high background that must be suppressed.  
Figure~\ref{excitation} shows the level excitation scheme.

\subsection{Experimental details}
\begin{figure}[tbp]
\includegraphics[width=0.3\textwidth]{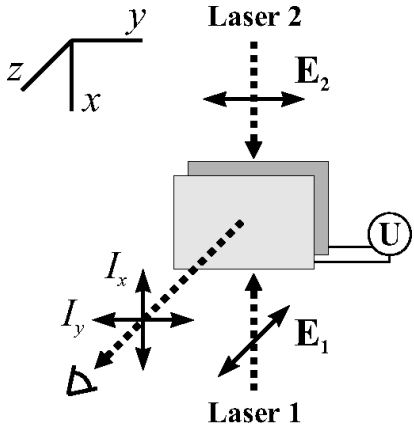}  
\caption{Experimental geometry.}
\label{exp}
\end{figure}

We studied cesium vapor at room temperature in a glass cell.  
The experimental setup is essentially the same as in Ref.~\cite{Auz06}.
We could apply an electric field between two transparent 
electrodes inside the cell, which were separated by a 2.5 mm gap.
Figure~\ref{exp} shows a schematic diagram of the experimental setup and
geometry.  The most crucial detail of the experiment is the
relative orientation of the electric field and the polarization vectors of
the linearly polarized laser radiation.  The first laser, which
excited the $6$S$_{1/2} \longrightarrow 6$P$_{3/2}$ transition, was polarized
with its polarization vector {\bf E$_1$} parallel to the dc electric field
{\bf $\mathcal{E}$}, which was along the $z$-axis.  The second laser, which
excited the 6P$_{3/2} \longrightarrow$ 7,9,10D$_{5/2}$ transition, was sent  
in a counter propagating direction and was polarized perpendicular to the
first, with polarization vector {\bf E$_2$}
parallel to the $y$-axis.  We observed the laser induced fluorescence
(LIF) at the $n$D$_{5/2} \longrightarrow$ 6P$_{3/2}$ transition along the
$z$-axis through the transparent electrodes.  A linear polarizer selected the
intensities of the LIF polarization components along the $x$ or $y$-axes $I_x$
or $I_y$.  Since the LIF was observed at the same frequency at which the second
laser was operating, it was necessary to suppress carefully the scattered
light by means of diaphragms.  The scattered light accounted for between 30\%
and 50\% of the measured signal.   We checked that this background remained stable
during the measurements and subtracted it from the signals.  The LIF passed
through an MDR-3 monochromator with 2.6~nm/mm inverse dispersion and was
recorded with a PMT in photon counting mode during one second time intervals.    

The first laser was always a diode laser (based on an LD-0850-100sm laser
diode) and was tuned to excite the 6$^2$P$_{3/2}$ state from the $F=3$ hfs component of the ground
state.  
We chose to
excite from the $F=3$ level, because in this way we could avoid the $F=6$
level of the $n$D$_{5/2}$ final state.  The $F=6$ level contained no level 
crossings and thus would contribute only background.  We took
advantage of a sideband of the radiation of the first laser in order to 
achieve broadband excitation. 

For the second excitation step, we used a
diode laser (based on a Hitachi HL6738MG laser diode) in the case of the 7D$_{5/2}$ state 
and a CR699-21 ring dye laser with Rhodamine 6G dye in the case of the 
9D$_{5/2}$ and 10D$_{5/2}$
states.  The second laser was operated in single mode regime.  
We recorded data at different values of the detuning of the second laser
in order to compare the results obtained at different detunings with simulations.
A HighFinesse WS/6 wavemeter allowed us to measure changes in the lasers'
detuning with a resolution of 30 MHz. 
However, in general we operated at the detuning that maximized the
fluorescence signal.  
When
the second laser was the diode laser, we jittered its output frequency over a
range of approximately 1.2 GHz by applying a sawtooth wave with a frequency of tens  of Hertz to a
piezoelectric crystal mounted to its feedback grating.  The laser power
was of the order of a few mW, and the laser beam diameters were approximately 
1 mm.  

The electric field produced in the cell was calibrated with measurements of
level-crossing signals for the 10D$_{3/2}$ state of cesium as in~\cite{Auz06}.  
The level-crossing
resonance positions obtained with our cell were compared with the crossing
points calculated from the tensor polarizability of~\cite{Xia97} and the hfs
constant $A$ of~\cite{Ari77}.  The overall uncertainty on the electric field
magnitude was estimated to be about 1\%.    

\subsection{Experimental results}

\begin{figure}[htbp]
\includegraphics[width=0.4\textwidth]{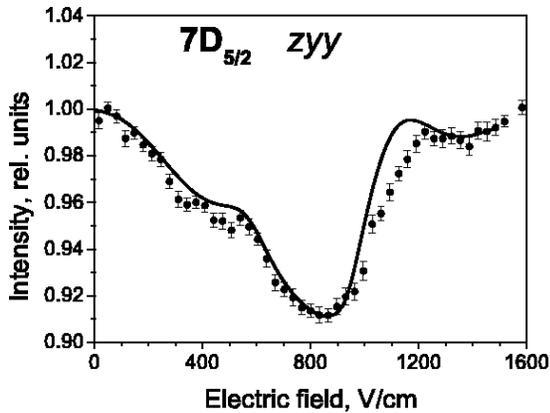}  
\caption{Experimental results for 7D$_{5/2}$.}
\label{res7}
\end{figure}

\begin{figure}[htbp]
\includegraphics[width=0.4\textwidth]{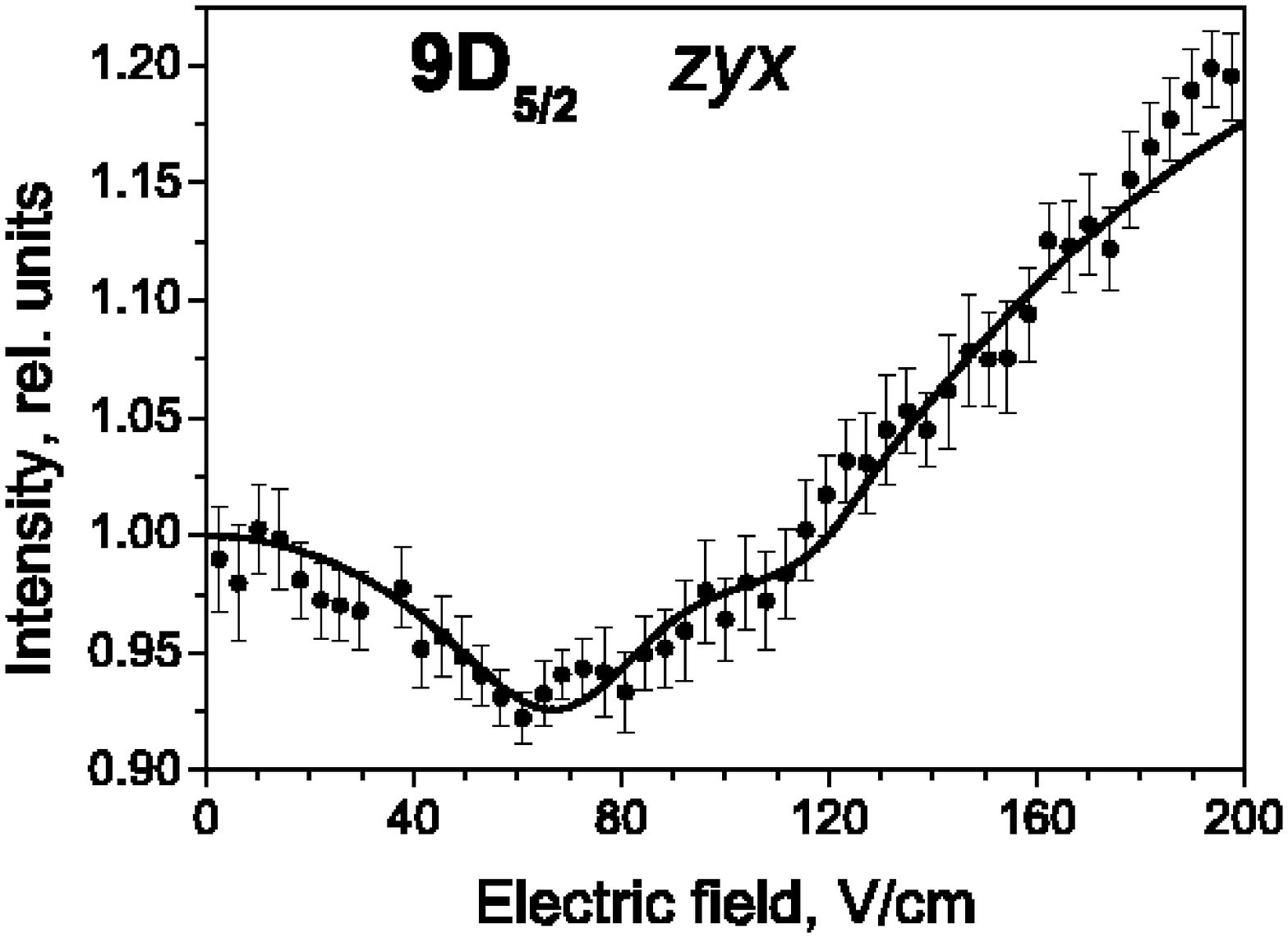} 
\includegraphics[width=0.4\textwidth]{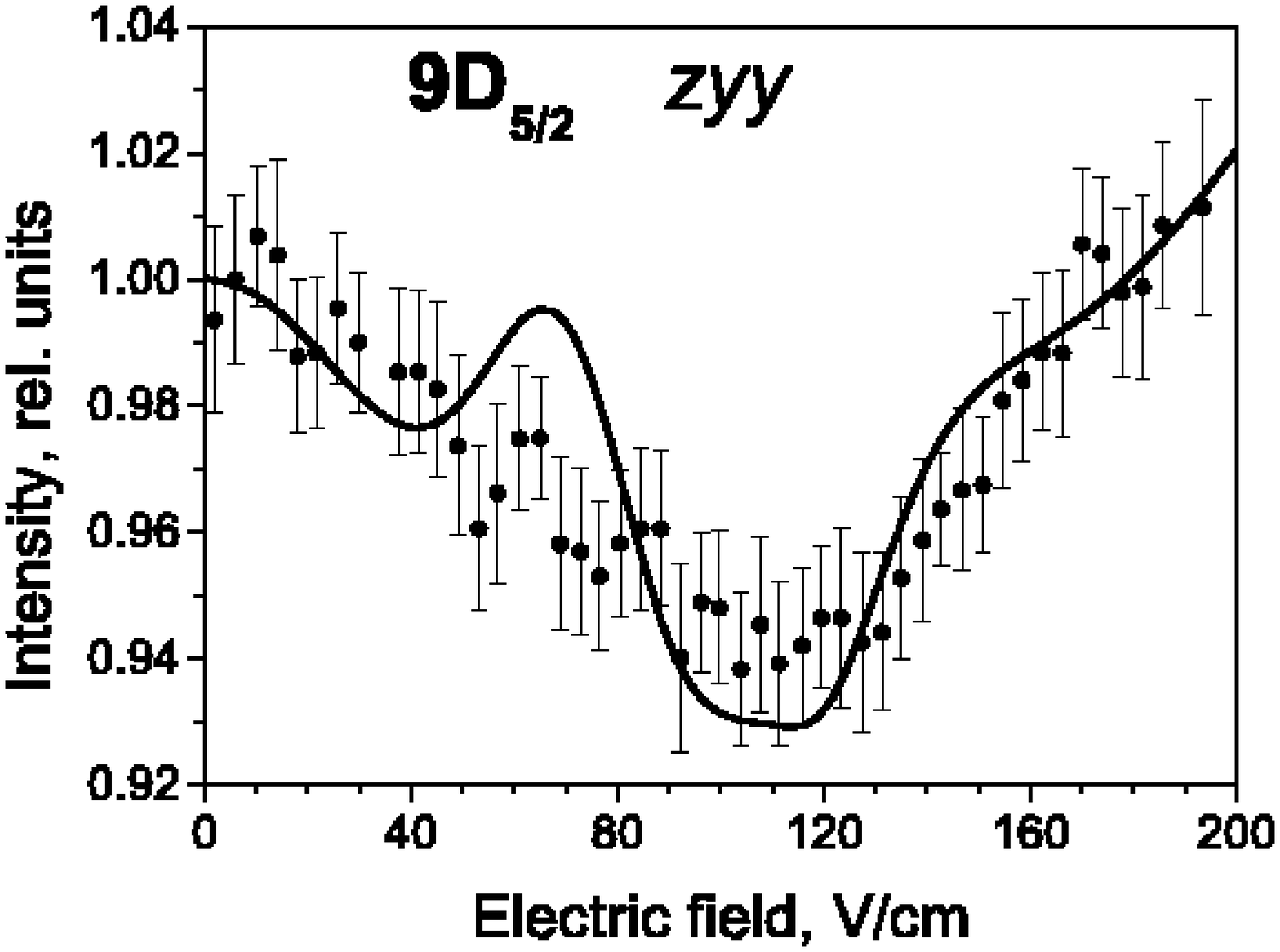} 
\caption{Experimental results for 9D$_{5/2}$.}
\label{res9}
\end{figure}

\begin{figure}[htbp]
\includegraphics[width=0.4\textwidth]{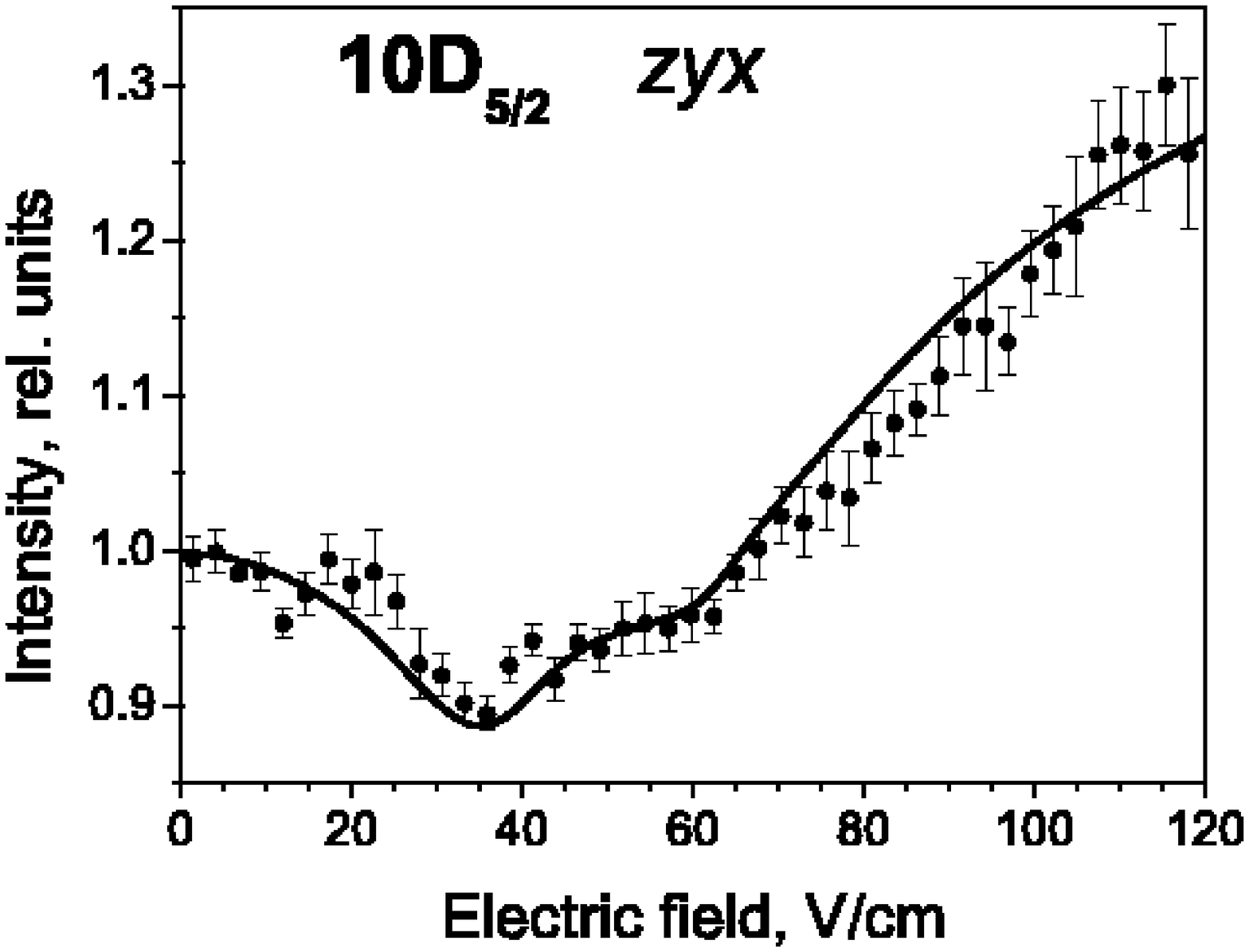} 
\includegraphics[width=0.4\textwidth]{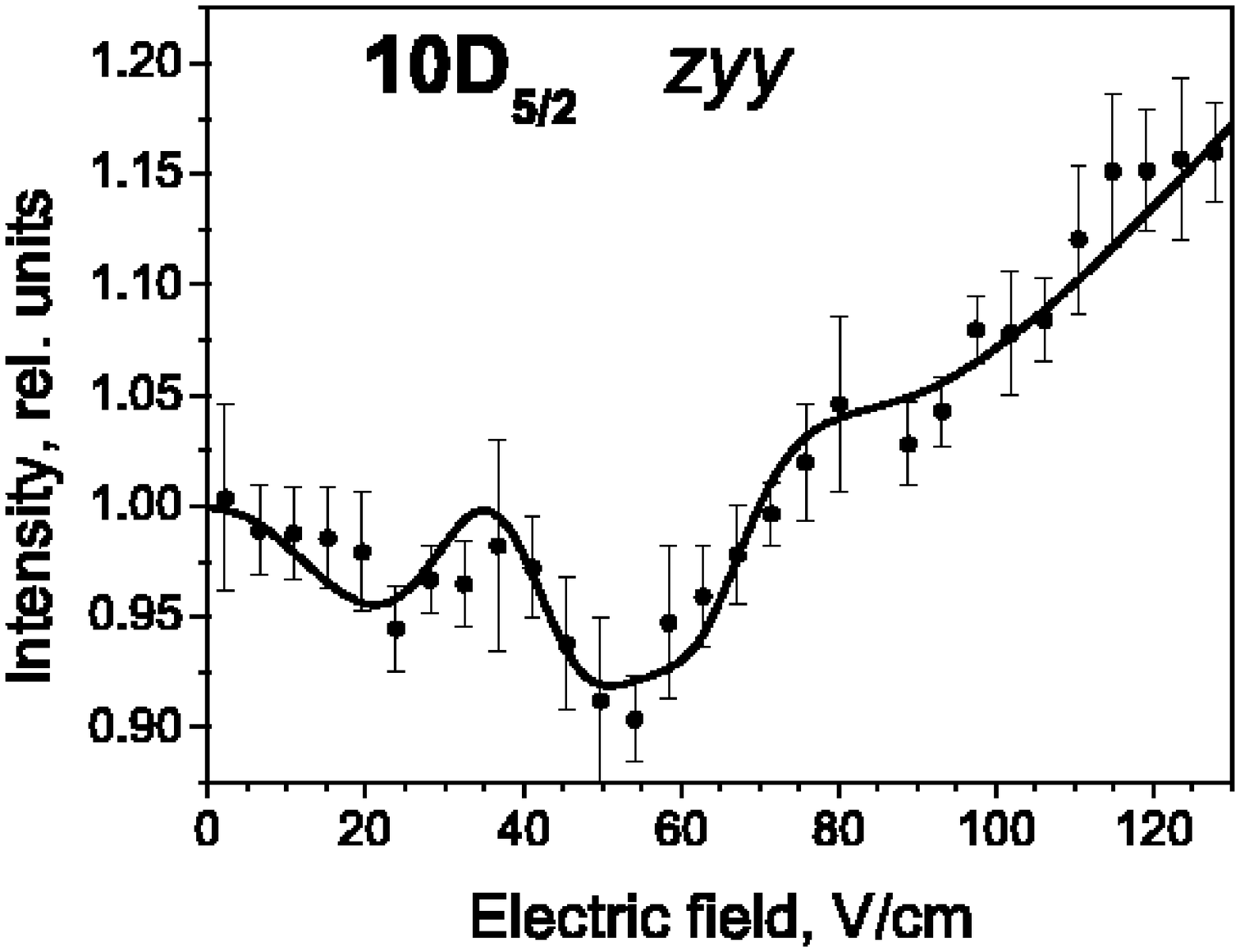} 
\caption{Experimental results for 10D$_{5/2}$.}
\label{res10}
\end{figure}

We plot with markers the measured LIF intensity as a function of the electric
field for the $n$D$_{5/2}$ states of cesium in 
Figures~\ref{res7}--\ref{res10}.  Signals for 
different experimental geometries are plotted.  
We label the experimental geometry as $zyy$ or $zyx$, where the first and the
second letters, $z$ and $y$, denote the orientation of the polarization of the
first and second lasers, $\mathbf{E_1} || z$ and $\mathbf{E_2} || y$, and
the third letter, $x$ or $y$, denotes the polarization direction of the observed
LIF.  
The solid line in the figure indicates the results of the
simulations that are described below.  As inputs to the theoretical
model, we used the tensor polarizabilities calculated with the relativistic 
many-body approach described in section III below.  


As can be seen from Figures~\ref{res7}, \ref{res9}, and~\ref{res10} there 
are no well-defined level crossing
resonances.  Nevertheless, a curve with multiple features is obtained, and these
features can be fitted with the results of a simulation based on a theoretical
model.
This simulation is described in the following subsection.  The fit involves
adjusting the hyperfine constant $A$ and those experimental parameters 
that we could not measure absolutely, such
as the laser detuning.  We fix the  
tensor polarizability $\alpha_2$ at the values that are obtained from the
calculations described in section III.

\subsection{Signal description}
Since well-defined resonances are no longer present in the signals of the
D$_{5/2}$ states, the data can be interpreted only by means of simulations
based on a detailed model.  Such a model was elaborated in detail  and
verified in  a previous publication~\cite{Auz06}, so we will only outline the
approach in what follows.

The model describes atoms that interact simultaneously with radiation produced
by two lasers with relatively broad spectral profiles, which  were necessary to 
excite coherently magnetic sublevels that
are split by an external electric field {\bf $\mathcal{E}$} (see
Fig.~\ref{lc}).  The model assumes that the atoms move classically and are
excited at the internal transitions.  Thus, the internal atomic dynamics can
be described by a semiclassical atomic density matrix $\rho$, which also
depends on the classical coordinates of the atomic center of mass.

The ground state of the Cs atom consists of two hyperfine levels with total
angular momentum $F_g=3$ and $F_g=4$, each containing $2F_g+1$ magnetic
sublevels.  The first laser excites the atoms from the ground state to the
6P$_{3/2}$ state, which contains hyperfine levels $F_e=2,3,4,$ and $5$.
The second laser excites the atom from the 6P$_{3/2}$ state to the
$n$D$_{5/2}$ state, which contained hyperfine levels $F_f=1,2,3,4,5,$ and $6$. 

The external electric field {\bf $\mathcal{E}$} partially decouples the
electronic angular momentum from the nuclear spin,
which implies that the magnetic sublevel energies no longer depend
quadratically on the electric field (see Fig.~\ref{lc}).  In order to obtain
the real dependence on the electric field, it is necessary to diagonalize
the full Hamilton matrix.  It is also necessary to take into consideration
that the decoupling of angular momentum from nuclear spin alters the dipole
transition probabilities between magnetic sublevels.  

The entire model is based on the Optical Bloch Equations (OBEs) for the
density matrix $\rho$ (see, for example, \cite{Ste84}) 
\begin{equation}
i\hbar \dfrac{\partial \rho }{\partial t}=\left[ \widehat{H},\rho
\right] +i\hbar \widehat{R}\rho.
\end{equation}%
The relaxation operator $\widehat{R}$ includes spontaneous emission and
transit relaxation.  We assume that the density of atoms is sufficiently low
that different velocity groups of thermally moving atoms do not interact.
The elements of the relaxation matrix are given in~\cite{Auz06}.  The
Hamiltonian $\widehat{H}=\widehat{H}_{hfs}+\widehat{V}$ includes the 
hyperfine Hamiltonian and the dipole interaction operator
$\widehat{V}=-\widehat{\mathbf{d}}\cdot \mathbf{E}\left( t\right)$,   
where $\widehat{\mathbf{d}}$ is the electric dipole operator and
$\mathbf{E}(t)$ is the electric field of the exciting radiation.  

The equations can be simplified by assuming that each laser excites only the
transition to which it is tuned.  We also apply the rotating wave
approximation for multilevel systems~\cite{Ari96} to the OBEs.  The resulting
stochastic differential equations can be further simplified by using the
decorrelation approach~\cite{Kam76}.  The stochasticity derives from the
random fluctuations of the laser radiation with finite spectral width.  This approach assumes that both lasers
are uncorrelated and that the integration time for each measurement is large
compared to the characteristic phase-fluctuation time of the exciting light
source.  The decorrelation approximation amounts to solving the equations of
the density matrix elements that correspond to optical coherences and taking a
formal statistical average over the fluctuating phases~\cite{Blush04}. This
procedure results in a system of equations that, when solved, yields the
observed signals. 

From the density matrix of the final state, one can obtain the fluorescence
intensities of a given polarization along the unit vector $\mathbf{e}$ from~\cite{Auz05,Coh61,Dya64}:%
\begin{equation}
I\left( \mathbf{e}\right) =\widetilde{I}_{0}\underset{g_{i},f_{i},f_{j}}{%
\sum }d_{g_{i}f_{j}}^{(ob)\ast }d_{e_{i}g_{i}}^{(ob)}\rho _{f_{i}f_{j}},
\label{Eq7.1}
\end{equation}%
where $\widetilde{I}_0$ is a constant and $d_{g_{i}f_{j}}^{(ob)}=$
$\left\langle g_{i}\left\vert \mathbf{d}%
\cdot \mathbf{e}\right\vert f_{j}\right\rangle $
is the matrix element between the ground and final states of the dipole
operator along a specific polarization direction $\mathbf{e}$, i.e., the $x-$
or $y-$ direction.

 \section{Calculation of scalar and tensor polarizabilities}
\subsection{Motivation}
The description of the signals obtained from the experiment described above 
depends on two atomic properties simultaneously: the hyperfine constant $A$ and
the tensor polarizability $\alpha_2$.  If one of these constants can be known
by independent means, the experiment provides a way to determine the other. 
In this section, we describe 
an all-order relativistic many-body calculation of the tensor polarizability
$\alpha_2$.  A reliable theoretical estimate of this constant, together with 
the experimental results of the previous section, can be used to estimate the
hyperfine constant $A$, which is difficult to calculate theoretically and
has not been measured to high precision for the 7,9, and 10D$_{5/2}$ states of cesium.

 \subsection{Method}

 \noindent     The scalar $\alpha_0$ and tensor $\alpha_2$ polarizabilities 
of an atomic state $v$ are calculated using formulas 
\begin{equation} \label{alpha0}
\alpha_0=\frac{2}{3(2j_v+1)}\sum_{n}
\frac{
\langle n\|D\| v \rangle^2 }
{E_{n}-E_{v}}
\end{equation}

\begin{eqnarray} 
\alpha_2&&=-4\left(\frac{5j_v(2j_v-1)}{6(j_v+1)(2j_v+1)(2j_v+3)}\right)^{1/2}\nonumber \\
&&\sum_{n}
(-1)^{j_v+j_n+1}\left\{
\begin{array}{ccc}
j_v & 1 & j_n \\
 1 & j_v & 2
\end{array}
\right\} 
\frac{
\langle n\|D\| v \rangle^2 }
{E_{n}-E_{v}},
\label{alpha2}
\end{eqnarray}
where $D$ is the dipole operator and the formula for $\alpha_0$ includes
 only the valence part of the polarizability. 
The contribution to $\alpha_0$ from the ionic core
 is negligible for the present calculation (16~$a^3_0$). 
 The sum over $n$ includes the $nP_{1/2}$, $nP_{3/2}$, and  $nF_{5/2}$ states
 for the calculation of the $D_{3/2}$ polarizabilities in cesium and the
 $nP_{3/2}$, $nF_{5/2}$, and  $nF_{7/2}$ states
 for the calculation of the $D_{5/2}$ polarizabilities.
 The sum over the intermediate states $n$ converges rather quickly and only the first few 
 terms need to be calculated accurately. Therefore, we separate the calculation of the polarizabilities
 into the calculation of the main term $\alpha^\textrm{main}$ and the evaluation of the remainder $\alpha^\textrm{tail}$. We include the contributions from the following states into the
 main term:  $6P$, $7P$, $8P$, $9P$, $10P$, $11P$, $12P$, $4F$, $5F$, $6F$, $7F$, and  $8F$
 to calculate the polarizabilities of all $D$ states considered in this paper. 
 We also include the contributions from the $9F$
 states into the calculation of the $\alpha^{\textrm{main}}(10D)$.
 All electric-dipole reduced matrix elements in Eqs.~(\ref{alpha0}, \ref{alpha2}) 
 that are needed for the calculation of the main term are calculated using the relativistic all-order 
 method, which is briefly described below. We use experimental energies from 
 \cite{NIST} in the main term calculations. 
 We note that the polarizabilities of the $9D$ and $10D$ states are very sensitive to the 
 values of the $9D-10P$ and $10D-11P$ energy differences, respectively, since 
 they are small ($50-100$~cm$^{-1}$).  We assume that the energies in Ref.~\cite{NIST}
 are accurate to all quoted digits. The remainders $\alpha^{\textrm{tail}}$ 
 are small for all sums and are calculated in the Dirac-Hartree-Fock (DHF) approximation.

 \begin{table*}
\caption{\label{tab7d1}  The contributions to
scalar and tensor polarizabilities for the $7D_{3/2}$ state in cesium. The 
corresponding energy differences
and the absolute values of the lowest-order (DHF) and final all-order electric-dipole
 reduced matrix elements are also listed. The energy differences are given in cm$^{-1}$.
 The electric-dipole  matrix elements  are given in atomic units ($ea_0$), and 
 the polarizabilities are given in 10$^3$~$a_0^3$, where $a_0$ is Bohr radius. }
\begin{ruledtabular}

\begin{tabular}{lrrrrrr}
\multicolumn{1}{c}{Contribution}&
\multicolumn{1}{c}{$nlj$}&
\multicolumn{1}{c}{$Z^{DHF}_{nlj,7D_{3/2}}$ }&
\multicolumn{1}{c}{$Z^{\textrm{SD}}_{nlj,7D_{3/2}}$ }&
 \multicolumn{1}{c}{$E_{nlj}-E_{7D_{3/2}}$ }&
\multicolumn{1}{c}{$\alpha_0(7D_{3/2})$} &
\multicolumn{1}{c}{$\alpha_2(7D_{3/2})$ \vspace{0.1cm}}\\
\hline \hline \\[-0.4pc]
  $\alpha^{\text{main}}(nP_{1/2})$  & $6P_{1/2}$ &  1.628  &    2.067  &  -14869.6   &    -0.011    &     0.011        \\
                                    & $7P_{1/2}$ &  4.030  &    6.580  &   -4282.2   &    -0.370    &     0.370        \\
                                    & $8P_{1/2}$ & 33.633  &   31.970  &    -338.7   &  -110.4(1.2) &   110.4(1.2)     \\
                                    & $9P_{1/2}$ & 13.535  &    8.734  &    1589.4   &     1.756    &    -1.756        \\
                                    & $10P_{1/2}$&  3.843  &    2.819  &    2679.2   &     0.109    &    -0.109        \\
                                    & $11P_{1/2}$&  2.026  &    1.537  &    3355.8   &     0.026    &    -0.026        \\
                                    & $12P_{1/2}$&  1.324  &    1.020  &    3805.0   &     0.010    &    -0.010        \\
 $\alpha^{\text{tail}}(nP_{1/2})$   &            &         &          &             &     0.041    &    -0.041 \\[0.5pc]
  $\alpha^{\text{main}}(nP_{3/2})$  & $6P_{3/2}$ &   0.794 &    0.983  &  -14315.5   &    -0.002    &    -0.002     \\
                                    & $7P_{3/2}$ &   2.111 &    3.336  &   -4101.2   &    -0.099    &    -0.079     \\
                                    & $8P_{3/2}$ &  15.190 &   14.351  &    -256.1   &   -29.4(3)   &   -23.5(3)    \\
                                    & $9P_{3/2}$ &   5.590 &    3.430  &    1634.1   &     0.263    &     0.211     \\
                                    & $10P_{3/2}$&   1.642 &    1.142  &    2706.1   &     0.018    &     0.014     \\
                                    & $11P_{3/2}$&   0.872 &    0.627  &    3373.2   &     0.004    &     0.003     \\
                                    & $12P_{3/2}$&   0.572 &    0.417  &    3816.9   &     0.002    &     0.001     \\
 $\alpha^{\text{tail}}(nP_{3/2})$   &            &         &           &             &     0.008    &     0.006 \\[0.5pc]
  $\alpha^{\text{main}}(nF_{5/2})$  & $4F_{5/2}$ &   9.165 &    13.027 &    -1575.4  &    -3.9(1)   &     0.79(3)      \\
                                    & $5F_{5/2}$ &  46.603 &    43.406 &      923.7  &    74.6(1.1) &   -14.9(2)     \\
                                    & $6F_{5/2}$ &   9.074 &     1.289 &     2281.9  &     0.027    &    -0.005      \\
                                    & $7F_{5/2}$ &   5.484 &     1.999 &     3100.4  &     0.047    &    -0.009      \\
                                    & $8F_{5/2}$ &   3.767 &     1.695 &     3631.1  &     0.029    &    -0.006      \\
 $ \alpha^{\text{tail}}(nF_{5/2})$  &            &         &           &             &     0.434    &  -0.087 \\[0.5pc]
  Total                             &            &         &           &             &   -66.8(1.6) &    71.2(1.2) \\
\end{tabular}
\end{ruledtabular}
\end{table*}

The all-order method used here sums infinite sets of many-body perturbation theory terms.
We refer the reader to Refs.~\cite{sd,cs,relsd} for a detailed
description of the approach. 
Briefly, the wave function of
the valence electron $v$ is represented as an expansion
\begin{eqnarray}
 |\Psi_v \rangle &= &\left[ 1 + \sum_{ma} \, \rho_{ma}
a^\dagger_m a_a + \frac{1}{2} \sum_{mnab} \rho_{mnab} a^\dagger_m
a^\dagger_n a_b a_a +
 \right. \nonumber \\
&+& \left. \sum_{m \neq v} \rho_{mv} a^\dagger_m a_v + \sum_{mna}
\rho_{mnva} a^\dagger_m a^\dagger_n a_a a_v  \right]|
\Phi_v\rangle  , \label{eq1}
\end{eqnarray}
where $\Phi_v$ is the lowest-order atomic state function, which is
taken to be the {\em frozen-core} Dirac-Hartree-Fock 
    wave function
 of a state $v$. This lowest-order atomic state function
can be written as $ |\Phi_v\rangle =a_v^{\dagger }|0_C\rangle,
$ where $|0_C\rangle $ represent DHF  wave function of a closed
core. The indices $m$ and $n$ designate excited states and indices $a$ and $b$ designate 
core states. The equations for the excitation coefficients are solved iteratively until 
the correlation energy converges to an acceptable accuracy. 
The excitation coefficients $\rho_{ma}$, $\rho_{mv}$, $\rho_{mnab}$, and $\rho_{mnva}$
are used to calculate the matrix elements, which can be expressed in the framework of the 
all-order method  as 
 linear or quadratic functions of the excitation coefficients. The electric-dipole 
 matrix elements as well as the hyperfine constants are calculated using the same approach.
 The expansion given by Eq.~(\ref{eq1}) is restricted to single and double (SD) excitations
 leading to the omission of certain fourth- and higher-order terms. 
 
 \begin{table*}
\caption{\label{tab7d2}  The contributions to the
scalar and tensor polarizabilities for the $7D_{5/2}$ state in cesium. The 
corresponding energy differences
and the absolute values of the lowest-order (DHF) and the final all-order electric-dipole
 reduced matrix elements are also listed. The energy differences are given in cm$^{-1}$.
 The electric-dipole  matrix elements  are given in atomic units ($ea_0$), and the
 polarizabilities are given in 10$^3$~$a_0^3$. }
\begin{ruledtabular}
\begin{tabular}{lrrrrrr}
\hline \\[-0.4pc]
\multicolumn{1}{c}{Contribution}&
\multicolumn{1}{c}{$nlj$}&
\multicolumn{1}{c}{$Z^{\textrm{DHF}}_{nlj,7D_{5/2}}$ }&
\multicolumn{1}{c}{$Z^{\textrm{SD}}_{nlj,7D_{5/2}}$ }&
 \multicolumn{1}{c}{$E_{nlj}-E_{7D_{5/2}}$ }&
\multicolumn{1}{c}{$\alpha_0(7D_{5/2})$} &
\multicolumn{1}{c}{$\alpha_2(7D_{5/2})$ \vspace{0.1cm}}\\
\hline \\[-0.4pc]
  $\alpha^{\text{main}}(nP_{3/2})$  & $6P_{1/2}$ &   2.375    &  2.909  & -14336.5  &  -0.014     &   0.014         \\
                                    & $7P_{1/2}$ &   6.303    &  9.679  &  -4122.2  &  -0.554     &   0.554         \\
                                    & $8P_{1/2}$ &  45.594    & 43.210  &   -277.1  & -164.3(1.7) & 164.3(1.7)      \\
                                    & $9P_{1/2}$ &  16.835    & 10.774  &   1613.1  &   1.755     &  -1.755         \\
                                    & $10P_{1/2}$&   4.939    &  3.555  &   2685.1  &   0.115     &  -0.115         \\
                                    & $11P_{1/2}$&   2.623    &  1.947  &   3352.3  &   0.028     &  -0.028         \\
                                    & $12P_{1/2}$&   1.720    &  1.294  &   3795.9  &   0.011     &  -0.011         \\
 $\alpha^{\text{tail}}(nP_{3/2})$   &            &            &         &           &   0.047     &  -0.047   \\[0.5pc]
  $\alpha^{\text{main}}(nF_{5/2})$  & $4F_{5/2}$ &   2.444    &  3.471  &  -1596.4  &  -0.184     &  -0.210        \\
                                    & $5F_{5/2}$ &  12.464    & 11.660  &    902.7  &   3.67(5)   &   4.20(5)      \\
                                    & $6F_{5/2}$ &   2.441    &  0.457  &   2260.9  &   0.002     &   0.003        \\
                                    & $7F_{5/2}$ &   1.472    &  0.590  &   3079.4  &   0.003     &   0.003        \\
                                    & $8F_{5/2}$ &   1.011    &  0.488  &   3610.1  &   0.002     &   0.002        \\
 $\alpha^{\text{tail}}(nF_{5/2})$   &            &            &         &           &   0.021     &   0.024   \\[0.5pc]
  $\alpha^{\text{main}}(nF_{7/2})$  & $4F_{7/2}$ &  10.925    & 15.292  &  -1596.5  &  -3.6(1)    &   1.28(4)    \\
                                    & $5F_{7/2}$ &  55.737    & 52.145  &    902.6  &  73.5(9)    & -26.2(3)       \\
                                    & $6F_{7/2}$ &  10.926    &  2.049  &   2260.8  &   0.045     &  -0.016        \\
                                    & $7F_{7/2}$ &   6.588    &  2.643  &   3079.3  &   0.055     &  -0.020        \\
                                    & $8F_{7/2}$ &   4.522    &  2.186  &   3610.1  &   0.032     &  -0.012        \\
 $\alpha^{\text{tail}}(nF_{7/2})$   &            &            &         &           &   0.416     &  -0.148   \\[0.5pc]
  Total                             &            &            &         &           &  -89.0(1.9) &    141.8(1.7)    \\
\end{tabular}
\end{ruledtabular}
\end{table*}

 \begin{table}
\caption{\label{tab9d}  The contributions to
scalar and tensor polarizabilities for the $9D_{3/2}$ and $9D_{5/2}$ states in cesium in 10$^3$~$a_0^3$.}
\begin{ruledtabular}
\begin{tabular}{lrr}
\multicolumn{1}{c}{Contribution}&
\multicolumn{1}{c}{$\alpha_0(9D_{3/2})$} &
\multicolumn{1}{c}{$\alpha_2(9D_{3/2})$ \vspace{0.1cm}}\\
\hline \hline \\[-0.4pc]
  $10P_{1/2}$     &  -1760(9) & 1760(9)       \\
  $10P_{3/2}$     &  -483(2)  & -386(2)    \\
  $6F_{5/2}$      &   -129(2) &  25.8(4)    \\
  $7F_{5/2}$      &    938(8) & -188(2)     \\
  Other          &     31    &  -22    \\
  Total           &  -1403(12)&1190(10)     \vspace{0.1cm} \\
\hline \\[-0.4pc]
\multicolumn{1}{c}{Contribution}&
\multicolumn{1}{c}{$\alpha_0(9D_{5/2})$} &
\multicolumn{1}{c}{$\alpha_2(9D_{5/2})$ \vspace{0.1cm}}\\
\hline \\[-0.4pc]
  $10P_{3/2}$&   -2653(12)&  2653(12)      \\
  $7F_{5/2}$ &    46.3(3)&   53.0(4)    \\
  $6F_{7/2}$ &    -117(2)&      41.9(6)  \\
  $7F_{7/2}$ &     927(6)&      -331(2)  \\
   Other     &      20   &       -30      \\
  Total       &  -1777(14)&   2386(13)    \\
\end{tabular}
\end{ruledtabular}
\end{table}

 We use B-splines \cite{Bspline} to generate a complete set of
DHF basis orbitals for the all-order calculation.  Here, we use  $N=70$ splines
for each angular momentum. The basis orbitals are constrained to a cavity of radius $R=220$~a.u. 
 The size of the cavity is taken to be large 
enough to fit all of the states
needed for the calculation of the main terms for all of the 
polarizabilities calculated in this work. 
The calculation of the polarizabilities of the $9D$ and $10D$ states requires such a large cavity 
since we need to be able to properly describe states up to 
$12P$ and $9F$. This work required extensive study of the numerical accuracy and
stability of the calculations. We verified that our basis set gives 
correct lowest-order (DHF)
values  for the energies of all relevant states and transition matrix elements
between these states. We have also verified that our basis set correctly reproduces 
DHF  values of the hyperfine constants for all the $nD_J$ states considered here. 
We find that it is necessary to use 70 splines to produce an accurate basis set. 
We also conducted an all-order calculation with a smaller cavity ($R = 90$~a.u.)
that  is appropriate for the calculation of the properties of the low-lying states and 
found that the properties of the low-lying states are accurately described by
our large $R=220$~a.u., $N=70$  basis set. Therefore, we conclude that numerically
accurate results can be obtained even for such highly-excited states as $12P$
with the use of large basis sets.

\subsection{Results}

The contributions to the scalar and tensor polarizabilities for the $7D_{3/2}$ state in cesium
are listed in Table~\ref{tab7d1}. We note that the calculation of the scalar and tensor polarizability differs
 only in the angular factor, and all matrix elements and energies are the same. 
 The corresponding energy differences
and the absolute values of the lowest-order and final all-order electric-dipole
 reduced matrix elements are also listed. 
 The energy differences are given in cm$^{-1}$.
 Electric-dipole  matrix elements  are given in atomic units ($ea_0$), and 
 polarizabilities are given in 10$^3$~$a_0^3$, where $a_0$ is the Bohr radius. The difference between the lowest-order values and the all-order values
 allows to evaluate the size of the correlation correction. The accuracy of our calculation is
 generally higher when the relative size of the correlation correction is smaller. 
 
 The contributions from all terms in $\alpha^{\textrm{main}}$ are listed separately
 to identify the most important terms. The remainder $\alpha^{\textrm{tail}}$ is separated to 
 $\alpha^{\textrm{tail}}(nP_{1/2})$, $\alpha^{\textrm{tail}}(nP_{3/2})$, 
 and $\alpha^{\textrm{tail}}(nF_{5/2})$ for the study of the convergence of these
 three sums. 

  We find that three contributions, from the $8P_{1/2}$, $8P_{3/2}$, and $5F_{5/2}$ states, are dominant.
Another term ($4F_{5/2}$) gives a small but significant contribution to the tensor
polarizability. 
Therefore, we conduct a more accurate calculation of the relevant matrix elements
and evaluate their uncertainties. The study of the breakdown of the correlation 
correction demonstrates that the main  contributions to these transitions
come from the terms containing only single valence excitation 
coefficients $\rho_{mv}$ (see Eq.~(\ref{eq1})). In such cases, it is possible to use 
a semi-empirical scaling procedure such as is described, for example, in Ref.~\cite{cs} to estimate
dominant classes of the omitted higher-order corrections. The single excitation coefficients
$\rho_{mv}$ are multiplied by the ratio of the experimental and theoretical 
correlation energy, and the calculation of the matrix elements is repeated
using  the modified excitation coefficients. The difference between the \textit{ab initio}
and  scaled SD all-order values for the particular matrix element is taken to be its uncertainty.
The relative uncertainty of the corresponding contribution to polarizability is 
twice the relative uncertainty of the matrix element. As we noted above, we   
assume that the experimental energies are accurate to all digits quoted in Ref.~\cite{NIST}.
The uncertainties of the total polarizability values are obtained by adding  
the  uncertainties of the  individual terms in quadrature.  The uncertainty in all remaining contributions is 
estimated to be  insignificant in comparison with the uncertainty of the dominant terms. 
\begin{table}
\caption{\label{tab10d}  Contributions to the
scalar and tensor polarizabilities  for the $10D_{3/2}$ and $10D_{5/2}$ states in cesium in 10$^3$~$a_0^3$. }
\begin{ruledtabular}
\begin{tabular}{lrr}
\multicolumn{1}{c}{Contribution}&
\multicolumn{1}{c}{$\alpha_0(10D_{3/2})$} &
\multicolumn{1}{c}{$\alpha_2(10D_{3/2})$ \vspace{0.1cm}}\\
\hline \hline \\[-0.4pc]
 $11P_{1/2}$& -4995(24)  &   4995(24)     \\
 $11P_{3/2}$& -1379(6)   &   -1103(5)  \\
 $7F_{5/2}$ &  -425(2)   &     85.1(4)    \\
 $8F_{5/2}$ & 2478(16)  &    -496(3)   \\
   Other    &    84   &     -65   \\
  Total     &    -4236(29)  &   3416(24)  \vspace{0.1cm} \\
\hline \\[-0.4pc]
\multicolumn{1}{c}{Contribution}&
\multicolumn{1}{c}{$\alpha_0(10D_{5/2})$} &
\multicolumn{1}{c}{$\alpha_2(10D_{5/2})$ \vspace{0.1cm}}\\
\hline \\[-0.4pc]
 $11P_{3/2}$& -7553(31)  &  7553(31)    \\
 $8F_{5/2}$ &   122(1)   &   140(1)   \\
 $7F_{7/2}$ &  -386(3)   &    138(1)   \\
 $8F_{7/2}$ &  2450(17)  &   -875(6)   \\
  Other          &    51  &   -89    \\
  Total     &   -5316(36) &   6867(32)    \\
\end{tabular}
\end{ruledtabular}
\end{table}
We observe significant cancellations between the dominant terms for both scalar and tensor
polarizabilities of the $7D_{3/2}$ state. However, the cancellation is more severe for 
the scalar polarizability, where the contributions from $8P_{1/2}$ and $5F_{5/2}$ states are comparable 
in size but have opposite signs. Therefore, we expect higher accuracy of our tensor polarizability calculation 
in comparison with the scalar one. 

 The contributions to scalar and tensor polarizabilities for the $7D_{5/2}$ state in cesium
are listed in Table~\ref{tab7d2}. The table is structured in exactly the same way as the one 
for the $7D_{3/2}$ state. We find that the contribution from the $8P_{3/2}$ state is clearly 
dominant and the cancellation is much less severe. For the tensor polarizability, 
the next largest term, $5F_{7/2}$, is six times as small as the dominant term. 
The accuracy of the matrix elements in the dominant terms is similar for the $7D_{3/2}$
and $7D_{5/2}$ states.
Therefore,
our calculation of the $7D_{5/2}$ polarizabilities is expected to be more accurate than that of the 
$7D_{3/2}$ polarizabilities.

\begin{table}
\caption{\label{comp0}  Comparison of the 
scalar polarizabilities  $\alpha_0$ for the $7D$, $9D$, and $10D$  states in cesium
with other theory and experiment. The 
 polarizabilities are given in 10$^3$~$a_0^3$. }
\begin{ruledtabular}
\begin{tabular}{lrrr}
\multicolumn{1}{c}{State}&
\multicolumn{1}{c}{This work} &
\multicolumn{1}{c}{Expt.} &
\multicolumn{1}{c}{Ref.~\protect{\cite{Wij94}}} \\
\hline \hline \\ 
 $ 7D_{3/2}$  &   -66.8(1.6)& -60(8) ~\protect{\cite{Wes87}}   & -65.2  \\
 $ 9D_{3/2}$  &   -1403(12) & -1450(120)~\protect{\cite{Fre77}} & -1400 \\
$ 10D_{3/2}$  &   -4236(29) & -4185(4) ~\protect{\cite{Xia97}}   & -4220  \\
$ 7D_{5/2}$  &   -89.0(1.9)& -76(8)  ~\protect{\cite{Wes87}}    & -87.1  \\
$  9D_{5/2}$  &   -1777(14) & -2050(100)~\protect{\cite{Fre77}} & -1770  \\
$ 10D_{5/2}$  &   -5316(36) & -5303(8)  ~\protect{\cite{Xia97}} & -5300 \\ 
\end{tabular}
\end{ruledtabular}
\end{table}

\begin{table}
\caption{\label{comp2}  Comparison of the 
tensor polarizabilities $\alpha_2$ for the $7D$, $9D$, and $10D$  states in cesium
with other theory and experiment. The 
 polarizabilities are given in 10$^3$~$a_0^3$. }
\begin{ruledtabular}
\begin{tabular}{lrrr}
\multicolumn{1}{c}{State}&
\multicolumn{1}{c}{This work}&
\multicolumn{1}{c}{Expt.}  &
\multicolumn{1}{c}{Ref.~\protect{\cite{Wij94}}} \\
\hline \hline \\[-0.4pc]
 $ 7D_{3/2}$  &    71.2(1.2) & 74.5(2.0)~\protect{\cite{Auz06}}   & 70.4 \\
              &              &   66(3)~\protect{\cite{Wes87}}     &       \\
  $ 9D_{3/2}$ &    1190(10)  & 1183(35)~\protect{\cite{Auz06}}   & 1190 \\
              &              &  1258(60) ~\protect{\cite{Fre77}}   &      \\
$ 10D_{3/2}$  &   3416(24)   &  3401(4)~\protect{\cite{Xia97}} & 3410  \\   
 $ 7D_{5/2}$  &    141.8(1.7)&     129(4)~\protect{\cite{Wes87}}    & 140   \\
$  9D_{5/2}$  &   2386(13)   &     2650(140) ~\protect{\cite{Fre77}}  &  2380  \\
$ 10D_{5/2}$  &   6867(32)   &    6815(20)   ~\protect{\cite{Xia97}}  & 6850  \\ 
\end{tabular}
\end{ruledtabular}
\end{table}

The contributions to
scalar and tensor polarizabilities  for the $9D_J$ and  $10D_J$ states in cesium 
are listed in Tables~\ref{tab9d} and ~\ref{tab10d}, respectively. The breakdown of the 
polarizability contributions is similar to that of the $7D$ polarizability calculations. 
We list only the 
dominant contributions separately and group all of the other contributions 
together in the rows labeled ``Other''. The uncertainty is evaluated using the method 
described above. The relative importance of the correlation corrections 
decreases with the principal quantum number $n$ and the cancellation of different
terms becomes less significant resulting in smaller uncertainties  of the polarizabilities 
for the $9D$ and $10D$ states in comparison with the uncertainties for the $7D$
states. Overall, the uncertainties of our polarizability calculation are $0.5\%-2.3\%$.

\subsection{Comparison with existing experimental values and other theory}

Our results for the scalar polarizabilities of the $7D_J$, $9D_J$, and $10D_J$ states in cesium 
are compared with the experimental values from 
Refs.~\cite{Wes87,Fre77,Xia97} and theoretical values from Ref.~\cite{Wij94} in Table~\ref{comp0}. 
The polarizabilities are given in 10$^3$~$a_0^3$. 
The conversion factor from the MHz/(kV/cm)$^2$ units 
to $10^3$ atomic units used in the present work is 
$10^{-7} h/(4 \pi \epsilon_0 a_0^3)=4.01878$, where $h$ is the Planck constant. 
The present values agree with the experimental results for $7D_{3/2}$, $9D_{3/2}$,
and $10D_{5/2}$ states within the corresponding uncertainties. There is some discrepancy with the 
accurate experimental value for the $10D_{3/2}$ state, but the discrepancy 
is only 1.5 of our estimated uncertainty. However, our values for the $7D_{5/2}$
and $9D_{5/2}$ states disagree significantly with the experimental values for these
states.  
The calculations for the 7D$_{5/2}$, 9D$_{5/2}$, and 10D$_{5/2}$ state 
polarizabilities are very similar.  Thus, the experimental values for the 
scalar polarizabilities are not consistent with each other according to our 
theoretical model.  Our calculations confirm the value for the 
10D$_{5/2}$ state to high precision, and one would have expected similar 
agreement in 
the case of the 7D$_{5/2}$ and 9D$_{5/2}$ state. 

The results for the tensor polarizabilities for the $7D_J$, $9D_J$, and $10D_J$ states in cesium 
are compared with the experimental values from 
Refs.~\cite{Wes87,Fre77,Xia97,Auz06} and theoretical values from Ref.~\cite{Wij94} in Table~\ref{comp2}. 
The polarizabilities are also given in 10$^3$~$a_0^3$. The present results for the $nD_{3/2}$
states support the measurements of Refs.~\cite{Xia97,Auz06} and disagree with the less precise
previous measurements \cite{Wes87,Fre77}. The comparison of the $nD_{5/2}$ values with 
experiment mirrors the result of the comparison for the scalar polarizabilities:
the $7D_{5/2}$ and $9D_{5/2}$ values differ significantly from the experiment
while the $10D_{5/2}$ value agrees with the precise experiment within the corresponding
uncertainties. Our values agree with the calculation of Ref.~\cite{Wij94} for all states
for both scalar and tensor polarizabilities.

\begin{table}
\caption{\label{corr}  The breakdown of the correlation correction to the 
 hyperfine constants $A$ for the $nD_{3/2}$ and $nD_{5/2}$ states in cesium calculated 
 using the SD all-order method. The expressions for all terms are  
given in \protect{\cite{sd}}. The values of the contributions for the dominant terms 
and total correlation correction are given in $\%$ relative to the lowest-order 
value for each state. The total contains contributions from all terms ($a-t$).
 The normalization factor is also listed.}
\begin{ruledtabular}
\begin{tabular}{lrrrrrr}
\multicolumn{1}{c}{Contribution}&
\multicolumn{1}{c}{$5D_{3/2}$}&
\multicolumn{1}{c}{$6D_{3/2}$}&
\multicolumn{1}{c}{$7D_{3/2}$}&
\multicolumn{1}{c}{$8D_{3/2}$}&
\multicolumn{1}{c}{$9D_{3/2}$}&
\multicolumn{1}{c}{$10D_{3/2}$}\\
\hline \hline
  Term a &    11\%  &  26\%   &  28\%  &   28\% &   28\% &   28\%  \\ 
  Term c &   127\%  &  57\%   &  36\%  &   28\% &   23\% &   21\%  \\ 
  Term d &    41\%  &   9\%   &   4\%  &    2\% &    2\% &    1\%  \\ 
  Term h &    13\%  &   9\%   &   5\%  &    4\% &    3\% &    3\%  \\ 
  Term p &    19\%  &  13\%   &  11\%  &   10\% &    9\% &    9\%  \\ 
  Total  &   214\%  & 118\%   &  87\%  &   75\% &   69\% &   65\%  \\ 
  Norm   &   1.10   & 1.14    & 1.12   &  1.10  &  1.10  &  1.09  \\ 
\hline
\multicolumn{1}{c}{Contribution}&
\multicolumn{1}{c}{$5D_{5/2}$}&
\multicolumn{1}{c}{$6D_{5/2}$}&
\multicolumn{1}{c}{$7D_{5/2}$}&
\multicolumn{1}{c}{$8D_{5/2}$}&
\multicolumn{1}{c}{$9D_{5/2}$}&
\multicolumn{1}{c}{$10D_{5/2}$}\\
\hline
  Term a  & -352\%  & -264\%& -228\%&   -213\% &   -205\%  &   -200\% \\
  Term c  &  120\%  &   54\%&   35\%&     27\% &     23\%  &     21\%  \\
  Term d  &   37\%  &    8\%&    3\%&      2\% &      2\%  &      1\%  \\
  Term h  & -154\%  &  -28\%&   -5\%&      3\% &      6\%  &      8\% \\ 
  Term n  &   18\%  &   16\%&   14\%&     13\% &     12\%  &     12\% \\ 
  Term p  &   13\%  &   10\%&    9\%&      8\% &      8\%  &      8\% \\ 
  Term r  &  -24\%  &  -18\%&  -15\%&    -14\% &    -14\%  &    -13\% \\ 
  Total   & -339\%  & -217\%& -184\%&   -171\% &   -164\%  &   -160\% \\ 
  Norm    &  1.09   &   1.12&  1.10 &   1.09   &    1.09   &    1.08 \\ 
\end{tabular}
\end{ruledtabular}
\end{table}

\section{Calculation of Hyperfine constants}
In this section we evaluate the current knowledge about the hyperfine constants
of the D$_{3/2}$ and D$_{5/2}$ states of cesium.   We describe a calculation of the hyperfine constants for the $5D_{3/2}-10D_{3/2}$ and
 $5D_{5/2}-10D_{5/2}$ 
states of $^{133}$Cs.  Then we compare the results of the calculation to previously measured values. 
 The calculation of the hyperfine constants also makes use of the relativistic all-order method and is done in the same way as the calculation of the 
electric-dipole matrix elements and with the same set of the excitation coefficients  
 $\rho_{ma}$, $\rho_{mv}$, $\rho_{mnab}$, and $\rho_{mnva}$ (see Eq.~(\ref{eq1})
).
 The breakdown of the correlation correction to the 
hyperfine constants $A$ for $nD_{3/2}$ and $nD_{5/2}$ states in cesium calculated 
 using the SD all-order method is given in Table~\ref{corr}. The expressions for the Terms $a, c, d, h, n,$ and $p$ are  
given in \protect{\cite{sd}}. These terms are linear or quadratic functions of the excitation
coefficients. The values of the contributions of the dominant terms 
and total correlation correction are given in $\%$ relative to the lowest-order 
value for each state. The normalization factor is also listed.
We find that the correlation correction is very large, especially for the $D_{5/2}$
states where it is several times as  large as  the lowest-order value and has an opposite sign.
Owing to such an enormous correlation correction, we do not expect our results 
to be very accurate for the $nD_{5/2}$ states. The scaling procedure described above 
or partial \textit{ab initio} inclusion of the triple excitation as described in Ref.~\cite{relsd}
can only evaluate corrections to terms $c$ and $d$, that are not dominant for any of the 
states except $5D_{3/2}$. Therefore, we can not make an accurate estimate 
of the uncertainty of our values that is independent from experimental measurements.

\begin{table}
\caption{\label{hyp}  The hyperfine constants $A$ (MHz) for the $nD_{3/2}$ and $nD_{5/2}$ states in cesium. The 
lowest-order, ``dressed'' third-order values, and all-order values are compared with previous experiments.
The experimental data are taken from ~\protect{\cite{Ari77}}. }
\begin{ruledtabular}
\begin{tabular}{lcccc}
\multicolumn{1}{c}{State}&
\multicolumn{1}{c}{DHF}&
\multicolumn{1}{c}{Third order}&
\multicolumn{1}{c}{All order}&
 \multicolumn{1}{c}{Expt.~\protect{\cite{Ari77}}} \\
\hline \hline
 $5D_{3/2}$  &    18.2  &     47.0   &    52.3  &  48.78(7)\\
 $6D_{3/2}$  &    9.27  &     21.5   &    17.8  &  16.30(15)\\
 $7D_{3/2}$  &   4.70   &     10.1       &    7.88  &   7.4(2)\\
 $8D_{3/2}$  &    2.65  &     5.46       &    4.20  &   3.94(8)\\
 $9D_{3/2}$  &    1.63  &     3.28       &    2.51  &   2.35(4)\\
$10D_{3/2}$  &    1.07  &     2.12       &    1.62  &   1.51(2)\\[0.5pc]
 $5D_{5/2}$  &    7.47  &     -32.3  &   -16.4  &  -21.24(8)\\
 $6D_{5/2}$  &    3.73  &     -8.15      &  -3.89   &   -3.6(10)\\
 $7D_{5/2}$  &    1.88  &     -2.67  &   -1.42  &   -1.7(2)\\
 $8D_{5/2}$  &    1.06  &     -1.15  &   -0.684 &   -0.85(20)\\
 $9D_{5/2}$  &    0.651 &     -0.592 &   -0.384 &   -0.45(10)\\
$10D_{5/2}$  &    0.428 &     -0.343 &   -0.238 &   -0.35(10)\\             
\end{tabular}
\end{ruledtabular}
\end{table}

\begin{table*}
\caption{\label{check}  The consistency check of the experimental hyperfine constants $A$ (MHz)
values for the $nD_{3/2}$ and $nD_{5/2}$ states in cesium. 
The actual experimental data from Ref.~\protect{\cite{Ari77}} are listed in the second column.
The columns labeled ``S($nD_{J}$)'', $n=6-10$, give data obtained by 
taking the experimental value for this particular $nD_{J}$ state and 
rescaling it for all the other states using the theoretical 
values for the correlation correction
as explained in the text. The uncertainty of the rescaled values comes only from the 
experimental uncertainty of the initial experimental value $nD_{J}$.}
\begin{ruledtabular}
\begin{tabular}{lrrrrrr}
\multicolumn{1}{c}{State}&
\multicolumn{1}{c}{Expt.~\protect{\cite{Ari77}}}&
\multicolumn{1}{c}{S($6D_{3/2}$)}&
\multicolumn{1}{c}{S($7D_{3/2}$)}&
\multicolumn{1}{c}{S($8D_{3/2}$)}&
\multicolumn{1}{c}{S($9D_{3/2}$)}&
\multicolumn{1}{c}{S($10D_{3/2}$)}\\
\hline \hline
$ 6D_{3/2}$ &  16.30(15) &         &   16.5(5)  & 16.3(4)  & 16.2(4) & 16.1(3)  \\  
$ 7D_{3/2}$ &   7.4(2)   &  7.33(6)&            &  7.35(16)&  7.3(1) & 7.25(12)  \\
$ 8D_{3/2}$ &   3.94(8)  &  3.93(3)&   4.0(1)   &          &  3.92(7)& 3.89(6)  \\  
$ 9D_{3/2}$ &   2.35(4)  &  2.36(2)&   2.38(6)  &  2.36(5) &         & 2.33(3)  \\ 
$10D_{3/2}$ &   1.51(2)  &  1.53(1)&   1.54(4)  &  1.53(3) & 1.52(3) &          \\
\hline
\multicolumn{1}{c}{State}&
  \multicolumn{1}{c}{Expt.}&
\multicolumn{1}{c}{S($6D_{5/2}$)}&
\multicolumn{1}{c}{S($7D_{5/2}$)}&
\multicolumn{1}{c}{S($8D_{5/2}$)}&
\multicolumn{1}{c}{S($9D_{5/2}$)}&
\multicolumn{1}{c}{S($10D_{5/2}$)}\\
\hline
 $ 6D_{5/2}$ &   -3.6(10) &             &  -4.5(5)  & -4.6(7)   &  -4.4(7)  & -5.2(1.1)  \\
 $ 7D_{5/2}$ &   -1.7(2)  &   -1.3(4)   &           & -1.7(4)   &  -1.6(3)  & -2.0(5)  \\    
 $ 8D_{5/2}$ &   -0.85(20)&   -0.6(2)   &  -0.83(10)&           &  -0.80(17)& -0.97(26) \\   
 $ 9D_{5/2}$ &   -0.45(10)&   -0.34(14) &  -0.47(6) & -0.48(11) &           & -0.56(16) \\   
 $10D_{5/2}$ &   -0.35(10)&   -0.21(9)  &  -0.29(4) & -0.30(8)  &  -0.28(6) &            \\
\end{tabular}
\end{ruledtabular}
\end{table*}

Our results for the hyperfine constants $A$ (MHz) for the $nD$ state in Cs are compared with previous experiments in Table~\ref{hyp}. We list the 
lowest-order and  ``dressed'' third-order values together with the SD all-order values.
The ``dressed'' third-order calculation has all lowest-order matrix elements replaced by 
``dressed'' matrix elements calculated in the random-phase approximation (RPA) \cite{WRJ}.
We find large discrepancies between the third-order and all-order results indicating very large
contributions from the fourth- and higher-order terms. Taking into account the very large 
size of the correlation correction and obviously large contributions from higher orders,
we find that the agreement of the all-order calculation with measured values is remarkably good. 

We have investigated the issue of the consistency of the experimental hyperfine
data using our calculation. 
Table~\ref{corr} demonstrates that 
 the breakdown of the correlation for the $6D-10D$ states is rather similar, especially for $nD_{3/2}$
 states. We note that $nD_{3/2}$ and $nD_{5/2}$ states have to be considered separately.
 The distributions of the correlation for both $5D_{3/2}$ and  $5D_{5/2}$ states are clearly
 very different from the ones for the other $nD$ states, and these states are omitted from the consistency check below. 
 For the $nD_{5/2}$ states, the relative contribution of Term $h$ changes sign; however, the contribution 
 from this term is small in comparison with the experimental uncertainty. To cross-check the experimental 
 data, we 
take the experimental value for one particular $nD_{J}$ state and 
rescale it for all the other states with the same $J$  using the theoretical 
values of the correlation corrections. 
The correlation correction is calculated as the difference between the final (experimental 
or theoretical) number and the lowest-order
DHF value.
For example, we take the experimental value $A^{\textrm{Expt}}_{6D_{3/2}}$ and determine 
 how much we need to scale our theoretical correlation correction for the $6D_{3/2}$ state
to obtain this value. The scaling factor is defined as 
$$
S(6D_{3/2}) =\frac{A^{\textrm{Expt}}_{6D_{3/2}}-A^{\textrm{DHF}}_{6D_{3/2}}} {A^{\textrm{SD}}_{6D_{3/2}}-A^{\textrm{DHF}}_{6D_{3/2}}},
$$
where $A^{\textrm{DHF}}$ and $A^{\textrm{SD}}$ are the lowest-order and all-order values
from Table~\ref{hyp} for the $6D_{3/2}$ state.
Next, we take our theoretical value for another state, for example, $7D_{3/2}$,
and rescale its correlation correction contribution using the scaling factor $S(6D_{3/2})$:
\begin{equation}
A(7D_{3/2})=S(6D_{3/2}) \times \left( A^{\textrm{SD}}_{7D_{3/2}}-A^{\textrm{DHF}}_{7D_{3/2}}\right) 
+A^{\textrm{DHF}}_{7D_{3/2}}.
\label{scale}
\end{equation}
Then, we calculate $A(8D_{3/2}), A(9D_{3/2})$, and $A(10D_{3/2})$ using Eq.~(\ref{scale}). We list these
values in the column labeled $S(6D_{3/2})$ of Table~\ref{check} which indicates that these values were 
obtained with the scaling factor $S(6D_{3/2})$.  We repeat the procedure using other $nD_{3/2}$
values to define the scaling factor.  
The uncertainty of the rescaled values comes only from the 
experimental uncertainty of the initial experimental value $A^{\textrm{Expt}}_{nD_{J}}$.
We find that all results in each row are consistent within the uncertainties,
leading to the conclusion that the experimental results are internally consistent. We 
note that such a procedure will not be able to detect a systematic shift of all the experimental results. 
Since we cannot accurately evaluate the uncertainty of the scaling procedure itself,
it is unclear if it can yield data that are more accurate than the corresponding experimental data,
even though some of the rescaled data has smaller uncertainties than the actual 
experimental data.
The accuracy of the rescaling is expected to be higher when $\Delta n$ between the original and scaled state
is the smallest.

\section{Analysis of Experimental Data and Estimate of the Hyperfine Constants}
        The theoretical calculations of the hyperfine constants described in the previous section as well as the experimental measurements
of~\cite{Ari77} contained large
uncertainties.  The scaling procedure seems to indicate that the experimental
values of the review~\cite{Ari77}, although taken from different sources, are consistent with each other.  Thus, there is an indication that the 
scaling procedure could yield slightly more accurate predictions of 
hyperfine constants of states in adjacent levels if the hyperfine constant of
one state is known.  The experiment described in Section II could provide an
independent cross-check of these findings.

With the tensor polarizabilities calculated in section III,  
the simulations described in
section II can be used to estimate the hyperfine constant.  First, we calculate
a series of simulated curves, varying those experimental parameters that we
cannot measure precisely, in particular the detuning of the lasers.  When the 
overall shape of the simulated curve matches the experiment, the positions of
the features depend on the values of the tensor polarizability $\alpha_2$ and
the hyperfine constant $A$.   
  
We assume that the tensor
polarizabilities calculated in section III for the 7,9, and 10D$_{5/2}$
states of cesium are the most accurate values available because of the
excellent agreement between the calculated and previously measured values
for the 10D$_{3/2}$ state of cesium.  By fixing the tensor
polarizability at the calculated value in our simulations, we can thus estimate
the hyperfine constant A from the level-crossing signals in
Figures~\ref{res7}, \ref{res9}, and \ref{res10}.

Table~\ref{constants}
summarizes the polarizabilities used in the simulations and the hyperfine
values $A$ obtained after a fit to the experimental data.    
\begin{table*}
\caption{\label{constants}  Comparison of the experimentally obtained hyperfine
  constants with previous experiments and the theory presented in this work}
\begin{ruledtabular}
\begin{tabular}{lcccc}
Cesium        & Calulated         & \multicolumn{3}{c}{hyperfine constant} \\
atomic        & tensor            & \multicolumn{3}{c}{(MHz)}\\
\cline{3-5}
state         & polarizability    & This work              & Previous                 &   Theory    \\ 
              & $(10^3 a_0^3)$    &                        & experiment               &             \\
\hline\hline
7D$_{5/2}$    & 141.8(1.7)        & -1.56(9)               & -1.7(2)~\cite{Ari77}     & -1.42       \\
9D$_{5/2}$    & 2386(13)          & -0.43(4)               & -0.45(10)~\cite{Ari77}   & -0.384      \\
10D$_{5/2}$   & 6867(32)          & -0.34(3)               & -0.35(10)~\cite{Ari77}   & -0.238      \\
\end{tabular}
\end{ruledtabular}
\end{table*}

Considering the difficulty in calculating the hfs constants, the results of
the relativistic 
many-body calculation for the hyperfine
constant $A$ agree reasonably well
with the experimental measurements for the 7D$_{5/2}$ and 9D$_{5/2}$ states
(within $\sim1.5\sigma$).  The large discrepancy in the case of the 
10D$_{5/2}$ state seems problematic, since the calculations should be 
internally consistent, if not completely reliable in absolute terms.  
This inconsistency could indicate that we slightly underestimated our
uncertainties.  
It is also
possible that the self-consistency check is less reliable in the case of the
$n$D$_{5/2}$ states, because the DHF term and the all order term differ even 
in their sign.

\section{Conclusion}
We obtained new values for the hfs constants $A$ of the 7,9, and 10D$_{5/2}$
states.  Our values agreed with previously measured values, but achieved
greater precision.  The values were
obtained by means of measured level-crossing signals, a detailed
theoretical description of these signals, and values for the tensor
polarizability calculated with an all-order relativistic many-body method.  We
demonstrated the all-order relativistic many-body method's reliability even in
highly excited states of $^{133}$Cs by comparing scalar and tensor
polarizabilities obtained by this method with previously experimentally
measured values for the 7,9,10D$_{3/2}$ and 7,9,10D$_{5/2}$ states of $^{133}$Cs.

Our calculated polarizability values were in good agreement with experiment 
except for the 7 and 9D$_{5/2}$ states.  However, the experimental values
reported for these states are called into question by the fact that values
reported in the same works for the 7D$_{3/2}$~\cite{Wes87} and
9D$_{3/2}$~\cite{Fre77} states also disagree with our calculations, whereas more
recent measurements of the 7 and 9D$_{3/2}$ states~\cite{Auz06} support our 
calculation, as well as previous calculations~\cite{Wij94}.  
The method was further applied to calculate values for the hyperfine constants 
$A$ in the 5-10D$_{3/2}$ and 5-10D$_{5/2}$ states.  Although the calculation
cannot be considered reliable in absolute terms, nevertheless they agreed
reasonably well in the case of the 7D$_{5/2}$ and 9D$_{5/2}$ states.  For the 
10D$_{5/2}$ state, the agreement was not as good.

\begin{acknowledgments}
We would like to thank Walter Johnson for providing his
``dressed'' third-order code for the evaluation of the 
importance of higher orders for the hyperfine constant calculation. We thank
Janis Alnis for help with the diode lasers and Robert Kalendarev for preparing
the cesium cells used in the experiment.
   The calculations of atomic properties were supported in part by DOE-NNSA/NV
Cooperative Agreement DE-FC08-01NV14050. The work of MSS was
supported in part by National Science Foundation Grant  No.\
PHY-04-57078.  The experimental measurements were supported by the NATO 
SfP~978029 Optical Field Mapping grant, Latvian National Research Programme
in Material Sciences Grant No. 1-23/50, and Latvian University Grant Y2-22AP02.
K.B., F.G., and A.J. gratefully acknowledge support from the European Social 
Fund. 
\end{acknowledgments}

\bibliographystyle{apsrev}
\bibliography{cs_lc}

\end{document}